\documentclass[aps,pre,twocolumn,amsmath,amssymb,showpacs,amsfonts]{revtex4}
\usepackage{epsfig}
\usepackage{graphicx}
\usepackage{dcolumn}
\usepackage{bm}

\begin{document}

\title{Evolution of collision numbers for a chaotic gas dynamics}

\author{Alexander Jonathan Vidgop$^1$}
\author{Itzhak Fouxon$^{2}$}

\affiliation{$^1$ Am haZikaron Institute, Tel Aviv 64951, Israel}
\affiliation{$^2$ Raymond and Beverly Sackler School of Physics and Astronomy,
Tel-Aviv University, Tel-Aviv 69978, Israel}
\begin{abstract}

We put forward a conjecture of recurrence for a gas of hard spheres that collide elastically in a finite volume.
The dynamics consists of a sequence of instantaneous binary collisions.
We 
study how the numbers of collisions of different pairs of particles grow as functions of time.
We observe that these numbers can be represented as a time-integral of a function on the phase space.
Assuming the results of the ergodic theory apply, we describe the evolution of the numbers by an effective Langevin dynamics. We use the facts that hold for these dynamics with probability one, in order to establish properties of a single trajectory of the system.
We find that for any triplet of particles there will be an infinite sequence of moments
of time, when the numbers of collisions of all three different pairs of the triplet will be equal. Moreover, any
value of difference of collision numbers of pairs in the triplet will repeat indefinitely. On the other hand, for 
larger number of pairs there is but a finite number of repetitions.
Thus the ergodic theory produces a limitation on the dynamics. 

\end{abstract}
 \maketitle


The systems of colliding hard-core particles take a special place in the theory of many-body systems. They were used extensively already by Boltzmann to study the fundamental principles of the statistical physics \cite{Boltzmann}. These systems are unique as they allow insight into the basic properties of complex systems that are usually postulated. Namely, rigorous facts on the ergodicity \cite{SinaiBook,Dorfman} of these systems are known. The famous Boltzmann-Sinai hypothesis states that systems of an arbitrary number $N\geq 2$ of elastic hard balls in a $d-$dimensional box with periodic boundary conditions (torus), $d\geq 2$, are ergodic in the phase space region where the trivial conserved quantities of the system are constant \cite{Sinai}. The exceptional feature of this hypothesis is that by today it can be considered as "almost proved", in contrast to the ergodicity hypothesis on other chaotic systems where the proof is generally absent. The first rigorous result was obtained in $1970$ by Sinai who proved that a system of two disks in a $2-$dimensional torus is ergodic \cite{Sinai2}. Notably, this result shows amply that the thermodynamic limit of a large number of particles is not necessary for the ergodicity. The extension of the proof to an arbitrary number of particles and arbitrary $d$ is "almost" complete by now, see e. g. \cite{Simanyi} and references therein, and also \cite{Szasz,Mulero}. As the system is equivalent to a billiard \cite{CM} - a single particle colliding elastically against the boundary of a certain manifold - we will call it below a "billiard".

In this Rapid Communication we use the ergodic theory not to deal with the equilibrium properties of the billiard, but rather to extract information on the structure of dynamics at finite times (this means considering non-equilibrium phenomena, as the equilibrium statistical physics, within the approach of the ergodic theory, describes the infinite-time averages).
The dynamics are a sequence of the events of binary collisions of particles. This sequence is an ordered list of pairs of particles, say $(1, 2)$ $(5, 6)$ $(7, 8)...$ meaning that first particles $1$ and $2$ collided, then $5$ and $6$, then $7$ and $8$ and so on.
Due to chaos this list looks like a random sequence of pairs. Here we find a deterministic constraint on that sequence.

The sequence of collisions $c_i$ is a discrete process taking $K(K-1)/2$ values corresponding to the pairs of the system with $K$ particles. This process is not a Bernoulli scheme, i. e. $c_i$ and $c_{i+1}$, determined by Newton's law, are not independent. However, the Bernoulli property, which is
likely to hold for the considered system \cite{GalOrn,Ornstein}, would suggest that $c_i$ and $c_{i+m}$ do become independent in the limit $m\to\infty$. Thus a coarse-graining of $c_i$ over a sufficient number of steps would produce a process which consecutive steps are already
independent to a good approximation, producing a random walk. Below we perform the analysis of the coarse-grained sequence in the formulation
that we found convenient.

We consider the numbers of collisions of pairs of particles up to a time $t$, cf \cite{Harald}. 
Roughly, for sufficiently large $t$ one can split the considered time-interval into sub-intervals with approximately independent numbers of collisions, so the total number of collisions is a sum of many i. i. d.
random variables and one can use the central limit theorem (CLT). Below we formulate this assumption using the ergodic theory. Then, the numbers of collisions of different pairs can be described using an effective system of Langevin equations. The sequence of collision events following from the Newtonian dynamics is statistically indistinguishable from a realization of the random Langevin dynamics, cf. \cite{Ornstein}. Here the statistics is defined by the volume of the initial conditions in the phase space corresponding to
the considered property.

The effective Langevin description allows to put forward a conjecture of recurrence that seems to be new. For any triplet of particles in the billiard, there is an infinite sequence of times at which the collision numbers of the corresponding three pairs of particles are equal. For $K=3$ this conjecture covers all the particles of the system. Furthermore, any prescribed value of two differences of collisions
will repeat infinitely many times. In contrast, a prescribed value of three and more differences, existing for $K>3$, repeats but a finite number of times. Significantly, the statements are dynamical and they hold for almost every trajectory of the system (i. e. with a possible exception of a set of trajectories the volume of the initial conditions of which is zero).

The analysis below applies to billiards both in $D=2$ and $D=3$ cases, where $D$ is the space dimension, with ramifications in $D=2$ case entailed by the slow, non-integrable decay of the correlation tails \cite{EW,PR}.
For definiteness one can think of hard balls of diameter $d$ that collide in a square cube with periodic boundary conditions. The dynamics is a succession of the discrete events of binary collisions. Starting from a given initial condition for particles' positions and velocities, one determines the time to the next collision and the pair that is going to collide. Pushing then the particles' positions and the velocities to the time after the collision, one iterates the procedure. We designate the number of collisions of particles $i$ and $j$ that occurred up to time $t$ by $N_{ij}(t)$. The important observation at the basis of the analysis below is that $N_{ij}(t)$ can be represented as an integral of a function on the phase space. To provide this representation we first consider a formal representation
of $N_{ij}(t)$ in terms of the Dirac $\delta-$function. We designate the collision times of the pair $i$ and $j$
by $t_{ij}^k$, and note that $r_{ij}^2(t)-d^2$ is a non-negative function of $t$ that vanishes only at $t=t_{ij}^k$. Introducing $F(\bm r, \bm v)\equiv 2 \delta\left[r^2-d^2\right]|\bm r\cdot \bm v|$, write formally
\begin{eqnarray}&&
\!\!\!\!\!\!\!\!\!N_{ij}(t)\!=\!\!\sum_k\!\!\int_0^t \!\delta\left[t'-t_{ij}^k\right]
dt'\!\!=\!\!\!\int_0^t\!\! F\left[\bm r_{ij}(t'),\! \bm v_{ij}(t')\right]
dt'
, \label{g1}
\end{eqnarray}
where $\bm r_{ij}(t)$ and ${\bm v}_{ij}(t)$ are the relative distance and velocity of particles $i$ and $j$, respectively. Using the above, one may
introduce a rigorous representation of $N_{ij}(t)$ by 
any family of smooth functions used to represent
the $\delta-$function. For example, using exponential profiles, and the fact that $r_{ij}^2(t)-d^2>0$ except at $t=t_{ij}^k$, we have
\begin{eqnarray}&&
\!\!\!\!\!\!\!\!\!\!\!\!\!\!N_{ij}(t)=\lim_{\epsilon\to 0}N^{\epsilon}_{ij}(t),\ \ N^{\epsilon}_{ij}(t)\equiv \int_0^t
\xi^{\epsilon}_{ij}(t')
dt',
\label{rigorous}
\end{eqnarray}
where $\xi^{\epsilon}_{ij}(t)$ are the regularized collision rates,
\begin{eqnarray}&&
\xi^{\epsilon}_{ij}(t)\equiv 2\epsilon^{-1}|\bm r_{ij}\cdot \bm v_{ij}|\exp\left[-\epsilon^{-1}\left(r_{ij}^2-d^2\right)\right],
\end{eqnarray}
rigorously. To see how the above works, single out a neighborhood of the collision $(t_{ij}^k-\delta, t_{ij}^k+\delta)$, where
$\delta$ is so small one can assume the particles $i$ and $j$ do not collide with other particles during that interval.
The quantity $\delta$ depends on properties of the system, like concentration. Considering a sufficiently small $\epsilon\ll \delta$ we have
\begin{eqnarray}&&
\!\!\!\!\!\!\!\!\!\int_{t_{ij}^k-\delta}^{t_{ij}^k+\delta} \xi^{\epsilon}_{ij}(t')
dt'
=\int_{t_{ij}^k-\delta}^{t_{ij}^k} dt'\frac{d}{dt'} e^{[d^2-r^2(t')]/\epsilon}
\nonumber\\&&\!\!\!\!\!\!\!\!\!\
-\int_{t_{ij}^k}^{t_{ij}^k+\delta} dt'\frac{d}{dt'}\exp\left[-\frac{r^2(t')-d^2}{\epsilon}\right]=1+o\left(\frac{\epsilon}{\delta}\right).
\end{eqnarray}
where we used that $\bm r_{ij}\cdot \bm v_{ij}<0$ before the collision and $\bm r_{ij}\cdot \bm v_{ij}>0$ after the collision. Note the latter
representation can be generalized straightforwardly to a dilute gas of particles with short-range interactions \cite{JI}.

The form (\ref{rigorous}) of $N_{ij}(t)$ suggests the use of the results of the ergodic theory. Ergodicity guarantees the existence of the average collision rate $\lim_{t\to\infty}N^{\epsilon}_{ij}(t)/t\equiv \nu^{\epsilon}$,
\begin{eqnarray}&&
\!\!\!\!\!\!\!\!\!\nu^{\epsilon}=\lim_{t\to\infty}\int_0^t  \xi^{\epsilon}_{ij}(t')\frac{dt'}{t}=\left
\langle \frac{2|\bm r\cdot \bm v|}{\epsilon}\exp\left[-\frac{r^2-d^2}{\epsilon}\right]\right\rangle
,\nonumber
\end{eqnarray}
where the angular brackets stand for the average over the microcanonical ensemble, and $\nu^{\epsilon}$ does
not depend on $ij$. It seems safe to assume that $\nu^{\epsilon}$ has a finite limit $\nu$ as $\epsilon\to 0$.
While the above applies to a single trajectory, here we address the behavior of the trajectories statistically. The statistics is defined by picking the initial condition at random in the allowed region of the phase space. The latter is defined by the trivial conserved
quantities of the system (for torus energy and momentum), i. e. we consider the microcanonical ensemble.

One expects that correlations of $\xi^{\epsilon}_{ij}$ at large times decay as $t^{-D/2}$, where $D$ is the space dimension \cite{EW,PR}. For $D=3$ the integral of
$t^{-D/2}$ converges and one may assume $\xi^{\epsilon}_{ij}(t)$ effectively has a finite correlation time $\tau^{\epsilon}_{cor}<\infty$, cf. e. g. \cite{Szasz}.
Here we make the main assumption underlying the conjecture proposed in this work. We assume the correlation time $\tau^{\epsilon}_{cor}$ has a finite
limit $\tau_{cor}$ at $\epsilon\to 0$. While this assumption seems very plausible (the representation seems to work for the dilute gas \cite{JI} where no
increase of the correlation time is known to us), it is this assumption that allows us to circumvent the singularity
in the representation (\ref{g1}).
Then at $t\gg \tau_{cor}$ the numbers $N^{\epsilon}_{ij}(t)$ are sums of roughly $t/\tau_{cor}\gg 1$ independent random variables and one can use the CLT, giving the following Gaussian approximation to the probability density function (PDF) $P(\{N_{ij}\}, t)$ of $N_{ij}(t)$ at $t\gg \tau_{cor}$:
\begin{eqnarray}&&
P(\{N_{ij}\}, t)=\frac{\exp\left[-(N_{ij}-\nu t)\Gamma^{-1}_{ij, mn}(N_{mn}-\nu t)/4t\right]}{\sqrt{(4\pi t)^{K(K-1)/2}\det \Gamma}},\nonumber
\end{eqnarray}
where 
the summation over repeated indices is assumed. The dispersion matrix $\Gamma_{ij, mn}$ describes the fluctuations of the collision rates $\xi_{ij}(t)$ and it is given by (the double brackets stand for the dispersion so that for any random variables $x$, $y$ we have $\langle\langle x y \rangle\rangle=\langle x y \rangle-\langle x \rangle\langle y \rangle$):
\begin{eqnarray}&&
\Gamma_{ij, mn}\equiv \int_0^{\infty} \langle\langle \xi_{ij}(0)\xi_{mn}(t)\rangle\rangle dt. \label{a}
\end{eqnarray}
The use of the CLT above neglects the tails in $P(\{N_{ij}\}, t)$. For certain billiards these tails can be even algebraic, see e. g. \cite{Sanders}. This is not a limitation for the analysis below, that concerns the probability of fixed values of $N_{ij}(t)-N_{mn}(t)$ at large $t$. At a sufficiently large $t$, any given value belongs to the bulk of the PDF of the differences and is describable by the CLT.

The simplest system for which the above relations apply is the $D=3$
system of $2$ balls in a 
torus. In this case we have but one pair of particles, so that the number of collisions $N$ that occurred by the time $t$ obeys
\begin{eqnarray}&&
P(N, t)=\frac{1}{\sqrt{4\pi \Gamma t}}\exp\left[-\frac{[N-\nu t]^2}{4\Gamma t}\right].
\end{eqnarray}
The above relation appears to be a fundamental result on a very basic system and it demands further studies, both theoretical and numerical that
we postpone for further work \cite{JI}. We note the same statistical distribution $P(\{N_{ij}\}, t)$ would result for the Langevin dynamics
\begin{eqnarray}&&
\frac{d N_{ij}}{dt}=\zeta_{ij}(t),\ \ \langle\zeta_{ij}\rangle=\nu,
\label{a1}\\&&
\langle \langle\zeta_{ij}(t)\zeta_{mn}(t')\rangle\rangle=2 \Gamma_{ij, mn}\delta(t-t'). \label{a2}
\end{eqnarray}
Thus the above stochastic dynamics gives the effective description of the collision numbers $N_{ij}(t)$ in quite the same sense as the usual
Langevin dynamics does: if one considers the dynamics over the temporal scale of coarse-graining that is much larger than $\tau_{cor}$ then
the two dynamics are statistically equivalent.

Eqs.~(\ref{a1})-(\ref{a2}) differ from similar results of the equilibrium
statistical physics. There the Langevin dynamics describes macroscopic quantities determined by a large number
of particles. Here the quantities $N_{ij}(t)$ are not macroscopic since the result holds even for the system of two particles (at least for
the torus). The macroscopic nature of the law is due to the consideration of the dynamics on a large time-scale.
While $N_{ij}(t)$ depend strongly on the details of the Newtonian mechanics at time-scales $\lesssim\tau_{cor}$, on a larger time-scale the dynamics forgets the details of the mechanism of collisions and $N_{ij}(t)$ are effectively Brownian motions with non-zero mean.

Using the above Langevin dynamics and facts that hold for the Brownian motion with probability one, we make deterministic predictions on the billiard. We note that the average rates of growth of $N_{ij}(t)$ are equal, so the differences ${\tilde N}_{ij}=N_{ij}-N_{12}$ are regular Brownian motions, 
\begin{eqnarray}&&
\frac{d {\tilde N}_{ij}}{dt}=\omega_{ij},\ \ \langle \omega_{ij}(t)\omega_{mn}(t')\rangle=2D_{ij, mn}\delta(t-t'),
\nonumber\\&&
D_{ij, mn}\equiv\Gamma_{ij, mn}+\Gamma_{12, 12}-\Gamma_{ij, 12}-\Gamma_{mn, 12}.
\end{eqnarray}
The non-diagonality of $D_{ij, mn}$ is not important for our considerations below. We note however that using that $D_{ij, mn}$ is symmetric it is always possible to pass to rotated ${\tilde N}_{ij}$ that perform independent Brownian motions.

We use the familiar fact that Brownian motion returns to the origin with unit probability for dimension lower or equal to the critical dimension $2$. In $D>2$ the return is probabilistic - there is a finite probability of return which is strictly less than one \cite{Montroll}, so the number of returns to the origin is always finite.
In the light of this, the differences ${\tilde N}_{ij}$, that exist for systems with $K\geq 3$, are seen to be special in the case $K=3$. Here the Brownian motion $({\tilde N}_{13}, {\tilde N}_{23})$ is two-dimensional. We arrive at the following conjecture of recurrence: for billiards with three particles there will almost always (i. e. with a possible exception of trajectories which initial conditions have zero volume) be a time when the numbers of collisions of all three pairs will equalize, $N_{12}(t)=N_{23}(t)=N_{12}(t)$.
Moreover, the system will be getting back to these equalized states an infinite number of times. Here it should be clear that the equality sign should be understood with a finite accuracy following both from the approximate nature of the Langevin equation and from the fact that the Brownian motion in $D=2$ is only neighborhood-recurrent and not point-recurrent as in $D=1$ \cite{MortersPeres}. This limitation is not important
qualitatively, since $N_{ij}(t)$ grow with time indefinitely, so the finite accuracy is irrelevant at large times. The result is quite distinct from the familiar Poincare recurrence theorem where the system gets back to the neighborhood of the same point in the phase space: the two recurrences are generally unrelated. Furthermore, based on the fact that the Brownian motion visits neighborhood of every point in the plane an indefinite number of times, the conjecture can be extended to the statement that every possible combination of $[N_{13}(t)-N_{12}(t), N_{23}(t)-N_{12}(t)]$ is going to occur and then recur an infinite number of times.

The above conjecture of recurrence does not hold for systems with $K\geq 4$. Here with a finite probability after $t=0$ there will never be again a situation where all $N_{ij}(t)$ are equal. The volume fraction of the initial conditions in the phase space for which all $N_{ij}(t)$ get equal at some time $t\gg \tau_{cor}$ is strictly less than one for $K\geq 4$ and it decreases as $K$ grows (for the explicit formula see \cite{Montroll}).
A limited version of the recurrence hypothesis holds, stating that for any three variables $N_{ij}$, $N_{kl}$ and $N_{mn}$ there is an infinite sequence of times at which any two differences of these numbers will take a fixed, preassigned value.
However, with probability one, three and more linearly independent differences of collision numbers, definable for a number of particles larger than three, will repeat but a finite number of times.
Finally, the return property 
in $D=1$, gives that the sequence of times at which 
$N_{ij}(t)=N_{mn}(t)$ is infinite for any $i$, $j$, $m$ and $n$.

We now rederive the conjecture qualitatively. We consider particles $1$, $2$ and $3$ and study the $D=2$ vector
$({\tilde N}_{13}, {\tilde N}_{23})$.
At large $t$ the probability that ${\tilde N}_{13}$ equals to zero decays as $t^{-1/2}$. Assuming effective independence of ${\tilde N}_{13}(t)$ and
${\tilde N}_{23}(t)$ we conclude that the probability that both latter functions equal zero decays as $t^{-1}$. Since the latter is non-integrable, the probability that there exists a finite time $t$ such that ${\tilde N}_{13}(t)={\tilde N}_{23}(t)=0$ is one. For three functions ${\tilde N}_{ij}(t)$ the corresponding probability would decay as $t^{-3/2}$ and the recurrence probability is strictly less than one. This argument
can be generalized to the recurrence of $({\tilde N}_{13}, {\tilde N}_{23})=(m, n)$ for any $m$, $n$.

The derivation can be generalized to $D=2$, where the usual Langevin dynamics does not hold. Such a generalization is important because
billiards on $D=2$ tori were studied extensively. For $D=2$ the correlations of $\xi_{ij}$ are expected \cite{EW,PR} to decay as $1/t$ making $\Gamma_{ij, mn}$ in Eq.~(\ref{a}) divergent. As a result the CLT cannot be used, however the recurrence still holds. The dispersion of ${\tilde N}_{ij}(t)$ grows at $t\ln t$, so the probability that ${\tilde N}_{ij}(t)=0$ decays as $(t\ln t)^{-1/2}$. The logarithmic correction does not change the convergence/divergence of the corresponding integrals and the argument above can be repeated. Thus we expect the conjecture also to hold in $D=2$.

We showed a use of the ergodic theory to approach a single realization of a chaotic dynamics. The basic observation is that under the assumption of a finite correlation time, natural for the considered systems, the PDF of the numbers of collisions of pairs of particles is described by the CLT. This allows to introduce Langevin's equations that provide an effective description of the dynamics of those numbers. Based on this description, the behavior of the realizations of which is well-known, one can draw conclusions on the behavior of the billiard's trajectories. For this one concentrates on facts that hold for the Langevin dynamics with probability one. Using the return properties of the random walks, we demonstrated that for every triplet of particles in the billiard there is an infinite sequence of times where
the differences of the collision numbers of all three pairs take an arbitrary fixed value. Furthermore, any value of three and more linearly independent differences of collision numbers, definable for $K>3$, repeats but a finite number of times. 
 
The conjecture involves the continuous approximation to the discrete process $c_i$ via the approximate dynamics of $N_{ij}(t)$. It would be advantageous to study the discrete random walk $c_i$ directly. This would demand the consideration of the one-dimensional random walk $N_{ij}(t)-N_{mn}(t)$ and the two-dimensional discrete random walk $[N_{ij}(t)-N_{mn}(t), N_{pr}(t)-N_{kl}(t)]$. Both walks occur on the lattice
with step size one. However, there are two major differences from the usual discrete random walk. First, the times between different steps of the
walk are random. Second, both, the time intervals between one step of the walk and the next one, and $c_i$ and $c_{i+1}$ are not independent. Thus the description of the appropriate random walk would demand either the consideration of detailed joint PDFs of time-intervals between the steps, or introducing a certain coarse-graining in time making the subsequent steps of the walk independent. The latter approach would involve a continuous approximation to $N_{ij}(t)$ and it can be said that our work does just that.

The described phenomena apply to any number of particles larger than one as long as the ergodic theory applies. Since they concern large numbers of collisions of the same pairs of particles, their practical observation (either numerical or experimental) demands considering systems with a relatively small number of particles. The results might also apply to the Lyapunov modes of large systems, that involve a small number of particles \cite{Harald}.

The hypothesis applies to almost every trajectory only. In particular, we do not consider those initial conditions for which the hard-sphere dynamics is ill-defined, which phase space volume was shown to vanish \cite{Alexander}.

Within the approximation used in the Boltzmann equation, the collisions in the gas of particles with short-range interactions are considered instantaneous. Thus the conjecture proposed here can be expected to hold for any gas of particles, where similar statistical properties of
$N_{ij}(t)$ hold. In particular, our derivation of the effective Langevin description, allows direct generalization to the gas [for the gas,
one can again introduce the time-scale $\delta$ where the colliding pair is "isolated" and the appropriate spatial scale $\sqrt{\epsilon}$].
We will report elsewhere the numerical studies of the conjecture proposed here, both for hard spheres and for the gas with short-range interactions \cite{JI}.

In this work we used the ergodic theory to derive quite detailed dynamical properties of almost all trajectories of the billiard.
We showed that any prescribed value of two differences of collision numbers will repeat indefinitely. On the other hand, any value 
of three and more
differences of collision numbers repeats but a finite number of times. These properties seem to be non-obvious from dynamical standpoint. The study of the underlying assumption of the asymptotic independence of $c_i$ and $c_{i+m}$ at large $m$,
both numerical and theoretical, is the subject for further analysis \cite{JI}.

We thank N. Chernov and N. Simanyi for very useful remarks that helped to improve the paper significantly.



\end{document}